# High-dimensional entanglement-enabled holography for quantum encryption

Ling-Jun Kong*, Yifan Sun*, Furong Zhang, Jingfeng Zhang, and Xiangdong Zhang†

*Key Laboratory of advanced optoelectronic quantum architecture and measurements of Ministry of Education, Beijing Key Laboratory of Nanophotonics & Ultrafine Optoelectronic Systems, School of Physics, Beijing Institute of Technology, 100081 Beijing, China.*

*\*These authors contributed equally to this work.*

*†Author to whom any correspondence should be addressed. E-mail: zhangxd@bit.edu.cn*

## Abstract

As an important imaging technique, holography has been realized with different physical dimensions of light, including polarization, wavelength, and time. Recently, quantum holography has been realized by utilizing polarization entangled state with the advantages of high robustness and enhanced spatial resolution, comparing with classical one. However, the polarization is only a two-dimensional degree of freedom, which greatly limits the capacity of quantum holography. Here, we propose a method to realize high-dimensional quantum holography by using high-dimensional orbital angular momentum (OAM) entanglement. A high capacity OAM-encoded quantum holographic system can be obtained by multiplexing a wide range of OAM-dependent holographic images. Proof-of-principle experiments with four- and six-dimensional OAM entangled states have been implemented and verify the feasibility of our idea. Our experimental results also demonstrate that the high-dimensional quantum holography shows a high robustness to classical noise. Furthermore, OAM-selective holographic scheme for quantum encryption is proposed and demonstrated. Comparing with the previous schemes, the level of security of holographic imaging encryption system can be greatly improved in our high-dimensional quantum holography.

The concept of holography was first introduced by Gabor [1,2], in 1948. Following the pioneering work of Gabor, Leith and Upatnieks applied the principle of holography to free-space optical beams with the advent of the laser [3], and Brown and Lohman invent the computer-generated hologram (CGH) [4]. Now, holography has been implemented with X-rays [5], electron beams [1], or neutron beams [6], and become an essential tool of modern optics of many applications for three-dimensional displays [7–10], microscopy [11,12], optical encryption [13,14], information storage [15], generation of topological structures [16–19] and so on. To date, the different physical dimensions of light, including wavelength [20–24], time [25,26] and polarization [27,28], have been utilized to carry independent information channels for increasing the capacity of holographic systems.

On the other hand, holography has been introduced into the quantum optics [29–32] and observed with polarization entangled state [33]. Information is encoded into the polarization degree of the entangled state, allowing us to image through dynamic phase disorder and even in the presence of strong classical noise with enhanced spatial resolution compared with classical coherent holographic systems. However, the current quantum holography is based on polarization, a two-dimensional degree of freedom. This greatly limits its channel capacity. If high-dimensional quantum holography can be realized, with keeping these advantages, the channel capacity will be further improved. Recently, orbital angular momentum (OAM) [34–39] has been introduced into the holographical system for encoding information. By using strong OAM selectivity in the spatial-frequency domain, high-dimensional OAM holography with high-capacity has been realized [40,41]. However, it is still a great challenge to introduce the strong OAM selectivity method into quantum holographic system, realize the high-dimensional OAM entanglement holography, and increase the channel capacity of the quantum holography.

In this work, we introduce the high-dimensional OAM entangled states into the holographic imaging and propose a method to realize the high-dimensional quantum holography. In our high-dimensional quantum holographic system, the OAM, represented by a helical wavefront with theoretically unbounded helical mode index, is implemented as an information carrier. By multiplexing of a wide range of OAM-dependent holographic images and using strong OAM selectivity method, we can obtain a high capacity OAM-encoded quantum holographic system. We have carried out the proof of principle experiment and verified the feasibility of the high-dimensional quantum holography and high capacity OAM-encoded quantum holographic system. The introduction of entanglement allows us to image even in the presence of strong classical noise. What's more, our results also show that the level of security of holographic imaging encryption system can be greatly improved by using high-dimensional quantum entangled state.

**Quantum OAM holography**—The arrangement of our high-dimensional quantum OAM holographic scheme is illustrated in Fig. 1(a). Photon pairs entangled in high-dimensional OAM are produced by pumping a β-barium borate (BBO) crystal under type I spontaneous parametric downconversion. The high-dimensional OAM entangled state of photon-pair is described as $|\Psi\rangle = \sum_{l=-\infty}^{l=+\infty} c_l |l\rangle_A |-l\rangle_B$. Here, $|l\rangle$ denotes a state of photon with an OAM of $l\hbar$, subscript A(B) labels the optical path-A(B), $c_l$ represents the amplitude distribution of $l$th-order OAM state. Experimental data for certifying high-dimensional OAM entangled source is offered in Sec. 1 of Supplementary Materials. Lenses L1 and L2 consist of a 4f system and image the plane of the BBO on the spatial light modulators, SLM-A and SLM-B. After interacting with the SLM-A, photon-A is coupled into a single mode fiber (SMF) with collecting lenses (Lc$_1$ and Lc$_2$) and detected by a single photon detector (D-A). Due to that only fundamental Gaussian mode state can be coupled into the SMF, the SLM-A and SMF are together used for projecting an OAM state (represented by $|x\rangle$) into $|0\rangle$. Therefore, the decoding operator of SLM-A can be described as $\hat{P}_A = |0\rangle_A \langle x|_A$. The SLM-B is used for displaying the holographic phase pattern and its function for photon-B is described by an operator $\hat{P}_B^H$. After interacting with the SLM-B and an optical Fourier transformation operated by lens Lf, the photon-B is collected by a multimode fiber (MMF) and detected by another single photon detector (D-B). The input end of the MMF scan in the Fourier plane of Lf. Thus, the holographic image with coincidence measurement between D-A and D-B is calculated as

$$H_i = \left| f\{ \hat{P}_B^H (\langle 0|_A \hat{P}_A |\Psi\rangle) \} \right|^2. \qquad (1)$$

Here, $f\{\cdot\}$ represents the Fourier transformation.

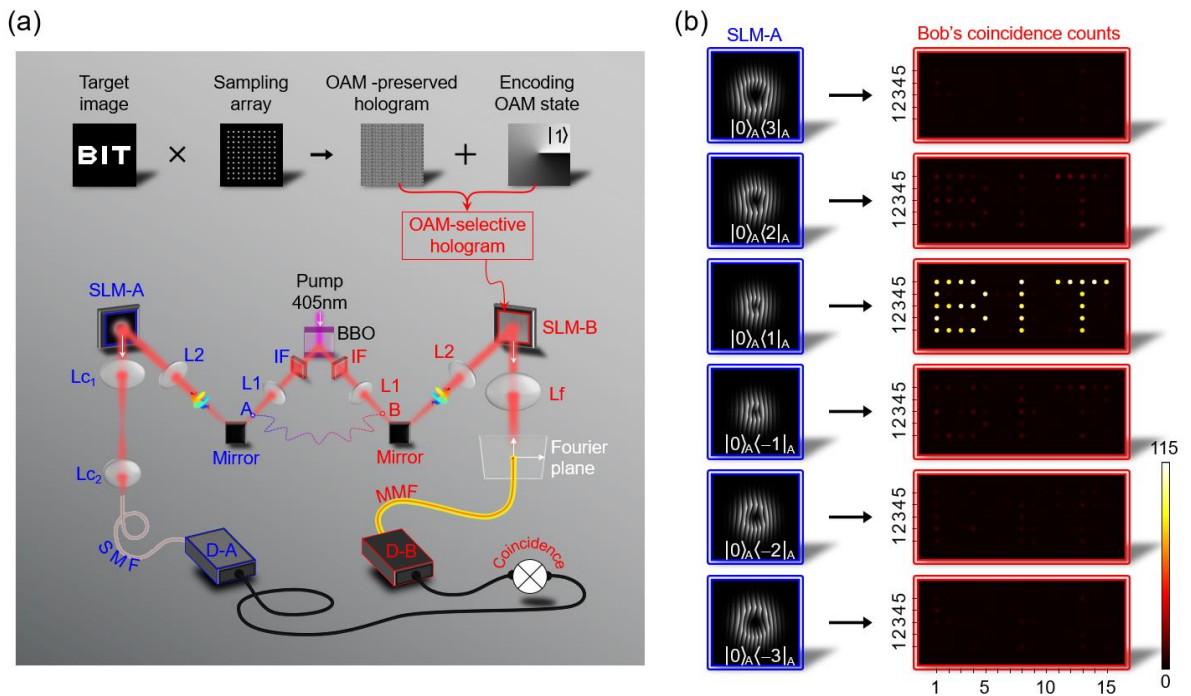

**FIG. 1. OAM entanglement-enabled quantum holography.** (a) The Schematic of setup. IF is an interference filter centred at 810 ± 3 nm. SLM-A (SLM-B) is the spatial light modulator. SMF and MMF represent single mode fiber and multi-mode fiber (with core diameter 25μm), respectively. D-A and D-B are two single photon detectors. Lc and Lf are two lenses. Top side shows the design of the OAM preserved and OAM-selective hologram (see Sec. 2 in Supplementary Materials for details). (b) Quantum OAM-selective holographic images reconstructed with coincidence measurement by implementing the OAM decoding operators $\hat{P}_A = |0\rangle_A\langle3|_A$, $|0\rangle_A\langle2|_A$, $|0\rangle_A\langle1|_A$, $|0\rangle_A\langle-1|_A$, $|0\rangle_A\langle-2|_A$, and $|0\rangle_A\langle-3|_A$. There are 15×5=75 pixels in each reconstructed holographic image. Coincidence measurement for each pixel takes 15s.

To realize such a quantum holography, the OAM-selective hologram should be displayed on the SLM-B. The OAM-selective hologram is obtained by encoding an OAM state $|l_e\rangle$ onto an OAM-preserved hologram, which is generated by multiplying a target image with a sampling array (see Sec. 2 in Supplementary Materials for details). An example is shown in the top side of Fig. 1(a), where the target image is the letters "BIT", the phase distribution of the OAM-preserved hologram is described with $\varphi$ and $l_e = 1$. In such a case, the function of the operator for photon-B is expressed as $\hat{P}_B^H = e^{j\varphi}|l_e + l\rangle_B\langle l|_B$. With setting the OAM decoding operator with $\hat{P}_A = |0\rangle_A\langle3|_A, |0\rangle_A\langle2|_A, |0\rangle_A\langle1|_A, |0\rangle_A\langle-1|_A, |0\rangle_A\langle-2|_A$, or $|0\rangle_A\langle-3|_A$ by displaying the corresponding phase patterns on the SLM-A, the quantum OAM-selective holographic images can be calculated by Eq. (1). Experiment has been carried out. The experimental results are given in Fig. 1(b), and the corresponding simulation results are shown in Sec.3 of Supplementary Materials. Due to that the encoded OAM state in OAM-selective hologram is $|1\rangle$, only the given OAM decoding operator $\hat{P}_A = |0\rangle_A\langle1|_A$ can convert each pixel of the holographic image into the Gaussian mode as $\hat{P}_B^H(\langle0|_A\hat{P}_A|\Psi\rangle) \propto \hat{P}_B^H|-1\rangle_B = e^{j\varphi}|0\rangle_B$. In the OAM-selective holographic experiment, only the Gaussian mode contributes to the reconstruction of holographic image (see Sec. 4 in Supplementary Materials for details). It is seen clearly that the OAM-selective holographic image can only be reconstructed for $\hat{P}_A = |0\rangle_A\langle1|_A$ with signal-to-noise ratio (SNR)~15.8. The agreements between experimental and theoretical results are very well, which means that the high-dimensional OAM entanglement holography has been realized.

***Quantum OAM-multiplexing holography.***—Since different OAM eigenstates are orthogonal to each other, the corresponding OAM-selective hologram can be designed according to each OAM state to encode the information independently. When these OAM-selective holograms are combined, OAM multiplexing holograms can be generated. Then, we can obtain a high capacity OAM-encoded quantum holographic system by multiplexing a wide range of OAM-dependent holographic images. The multiplexing approach is illustrated in Fig. 2(a). There are a series of target images, letters "O", …, "A", …, "M", …. The OAM-preserved hologram of each target image is generated by multiplying it with a sampling array. Then, the OAM-selective holograms are obtained by encoding OAM states $|l_1\rangle$, ..., $|l_i\rangle$, ..., $|l_j\rangle$, ... onto the OAM-preserved holograms. Finally, the OAM-multiplexing hologram is obtained by combine all OAM-selective holograms. Due to that the property of OAM states will be well maintained in each pixel of the reconstructed holographic images, incident OAM states can be used to reconstruct different holographic images from an OAM-multiplexing hologram. Thus, different target images can be reconstructed by using the OAM decoding operator with $\hat{P}_A = |0\rangle_A\langle l_1|_A$, ..., $|0\rangle_A\langle l_i|_A$, ..., $|0\rangle_A\langle l_j|_A$, ... respectively. Then, quantum holography with high capacity has been realized. The proof-of-principle experiment with six-dimensional OAM entangled states has been carried out. Here six target images are letters "O", "A", "M", "E", "Q", and "H". Six encoding OAM states are $|3\rangle$, $|2\rangle$, $|1\rangle$, $|-1\rangle$, $|-2\rangle$, and $|-3\rangle$. The experimental results are shown in Fig. 2(c). The images of letters "O", "A", "M", "E", "Q", and "H" can be obtained by using the OAM decoding operator of $\hat{P}_A = |0\rangle_A\langle 3|_A$, $|0\rangle_A\langle 2|_A$, $|0\rangle_A\langle 1|_A$, $|0\rangle_A\langle -1|_A$, $|0\rangle_A\langle -2|_A$, and $|0\rangle_A\langle -3|_A$, with SNR~8.8, 9.9, 10.4, 9.3, 9.4, and 6.8 respectively. The smaller SNRs for letters "O" and "H" is due to the smaller coincidence count rates of $|3\rangle_A|-3\rangle_B$ and $|-3\rangle_A|3\rangle_B$ (See experimental data offered in Sec. 1 of Supplementary Materials). On the other hand, the simulation results for ten-dimensional condition can be found in Sec. 4.4 of Supplementary Materials. All these results prove the feasibility of quantum OAM-multiplexing holography.

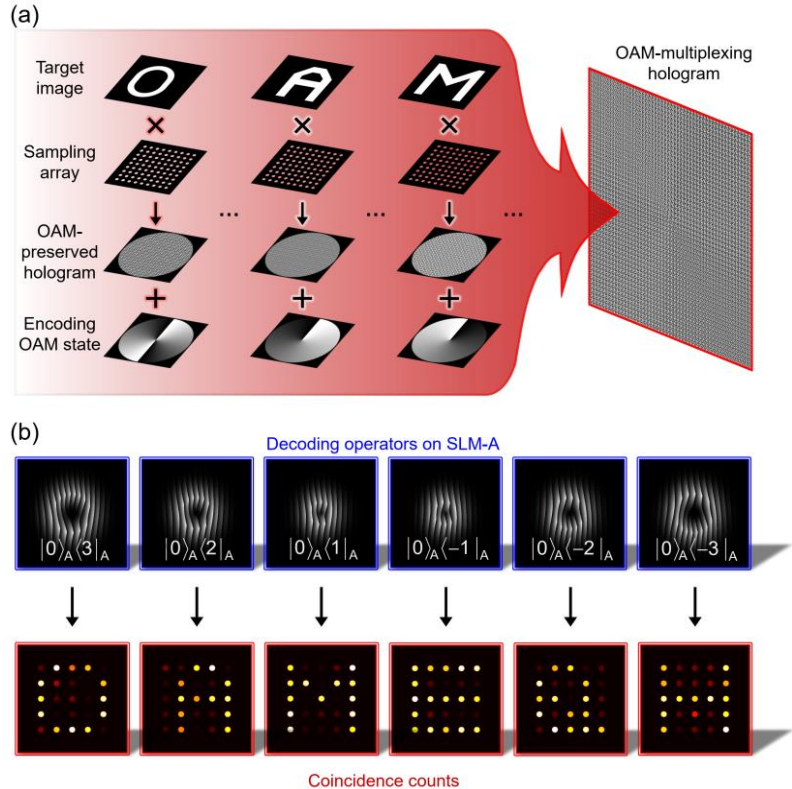

**FIG. 2.** Principle of quantum OAM-multiplexing holography. (a) Generation of an OAM-multiplexing hologram. (c) Experimental reconstructions of holographic images based on the OAM projective measurement operators $\hat{P}_A = |0\rangle_A\langle 3|_A$, $|0\rangle_A\langle 2|_A$, $|0\rangle_A\langle 1|_A$, $|0\rangle_A\langle -1|_A$, $|0\rangle_A\langle -2|_A$, and $|0\rangle_A\langle -3|_A$. There are 5 × 5 = 25 pixels and the coincidence measurement of each pixel takes 10s.

***Robustness of quantum OAM holography.***—One of the advantages of quantum holography is its robustness to classical noise, as demonstrated for amplitude objects illuminated by entangled pairs corrupted by static stray light [42,43]. To test such an effect, we illuminate white light of a lamp as classical noise in the experiment of quantum OAM holography (see Sec. 5 in Supplementary Materials for details). The experimental results for reconstructing the OAM-selective holographic images are given in Fig. 3(a). It is shown clearly that the encoded image (letters "BIT") can be accurately reconstructed for $\hat{P}_A = |0\rangle_A\langle 1|_A$ in the presence of classical stray light, with only a lower SNR compared to the case without stray light as shown in Fig. 1(b). In contrast, when we use single photon, instead of entangled photons, to reconstruct the holographic image (see Sec. 5 in Supplementary Materials for details), the experimental results are provided in Fig. 3(b). The images become too vague to tell the letters, no matter what kind of OAM state is used. In the classical OAM holographic system, the counts of coincidence measurement are related to the counts of each detector (D-A and D-B) and their time correlation. The classical noise greatly increases the counts of D-A and D-B, but slightly increases the counts of coincidence measurements because the counts coming from the classical noise do not have time correlation. The phenomena are similar to the case for the polarization entanglement-enabled quantum holography [33].

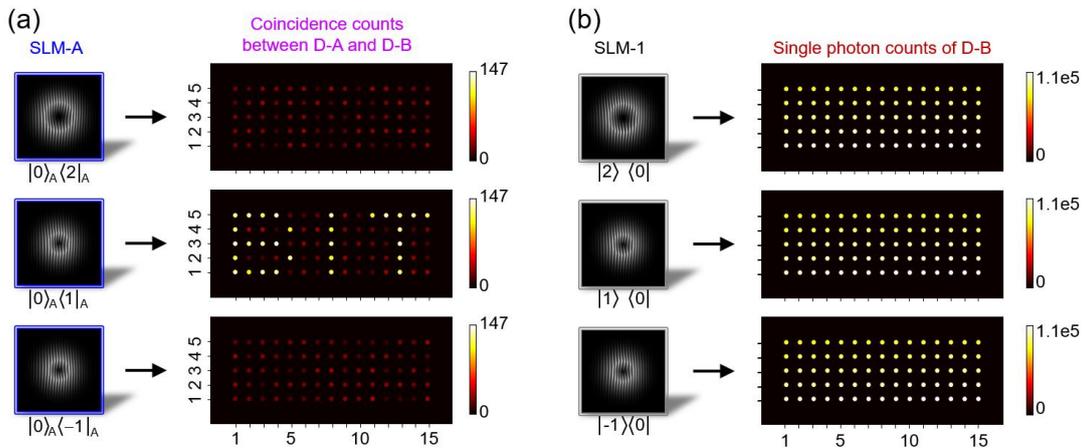

**FIG. 3.** Quantum OAM-selective holography in the presence of classical noise. (a) OAM-selective holographic images measured with quantum OAM holographic systems. (b) OAM-selective holographic images measured with classical OAM holographic systems.

**Entanglement-enabled holography for higher level security of holographic imaging encryption.**—What's exciting is that the quantum OAM-selective holography enables an imaging encryption strategy with a higher level of security compared with that based on classical one [40]. As is well known, a classical bit would have to be in state "0" or "1". However, a qubit is allowed to be in a coherent superposition of its basis state, such as $|0\rangle$ and $|1\rangle$. Therefore, if one employs a superposition state as the key to encrypt a text, searching for the key in the corresponding state space (Hilbert space) would be much harder than that for a key encoded by bits in its finite state-space. Here, we take the OAM states $|1\rangle$ and $|-1\rangle$ as an example. In the first-order OAM Poincaré sphere as shown in Fig. 4(a), $|1\rangle$ and $|-1\rangle$ are located at north and south poles, respectively [44,45]. Any state of superposition of $|1\rangle$ and $|-1\rangle$ can be described with a point on the Poincaré sphere, which is located by two angle parameters $\theta \in [-\pi/4, \pi/4]$ and $\phi \in [0, \pi]$. The points on the equator, for instance, represents the superposition of $|1\rangle$ and $|-1\rangle$ with equal probabilities; the relative phase of the superposition determines the orientation of the dark lines produced by coherent cancellation. In our quantum OAM holographic system, the information can be encoded with any point on the Poincaré sphere in principle, which implies that one has to guess the key among infinite number of possible states if no prior knowledge of the encryption is available. For high-dimensional entangled states, the situation is similar. More discussion has been offered in Sec. 6 in Supplementary Materials.

To verify the feasibility of entanglement-enabled holography for quantum imaging encryption, the experiments based on 4-dimensional OAM entangled states have been done. As shown in Fig. 4(b), four OAM-preserved holograms are obtained by multiplying four target images (letters 'O', 'A', 'M', and 'H') with a sampling array, respectively. Then, four OAM superposition states, $|S_e\rangle_1 = (|-2\rangle + |-1\rangle + |2\rangle + |1\rangle)/2$, $|S_e\rangle_2 = (|-2\rangle + j|-1\rangle - |2\rangle - j|1\rangle)/2$, $|S_e\rangle_3 = (|-2\rangle - |-1\rangle + |2\rangle - |1\rangle)/2$, and $|S_e\rangle_4 = (|-2\rangle - j|-1\rangle - |2\rangle + j|1\rangle)/2$, are encoded into the OAM-preserved holograms for generating four OAM-selective holograms, respectively. The OAM-multiplexing hologram is achieved by combining these OAM-selective holograms and displays on the SLM-B. Figure 4(c) displays the experimental results of OAM-selective holographic images for 64 kinds of decoding operators. The decoding operators are implemented by the SLM-A, which are described as $\hat{P}_A = |0\rangle_A [e^{-j\pi(\alpha_1-1)/2}\langle-2|_A + e^{-j\pi(\alpha_2-1)/2}\langle-1|_A + e^{-j\pi(\alpha_3-1)/2}\langle 2|_A + e^{-j\pi(\alpha_4-1)/2}\langle 1|_A]/2$. The values of parameters ($\alpha_1$, $\alpha_2$, $\alpha_3$, $\alpha_4$) are listed on each reconstructed OAM-selective holographic image. The images of letters 'O', 'A', 'M', and 'H' can be decoded with ($\alpha_1$, $\alpha_2$, $\alpha_3$, $\alpha_4$) = (1, 1, 1, 1), (1, 2, 3, 4), (1, 3, 1, 3), and (1, 4, 3, 2) (marked with the frames of purple line) with SNR = 11.3, 9.4, 11.5, and 10.8, respectively. These results demonstrate that only when one knows the exact expression of the superposition state used for imaging encryption, can he obtain the information of the encrypted four letters.

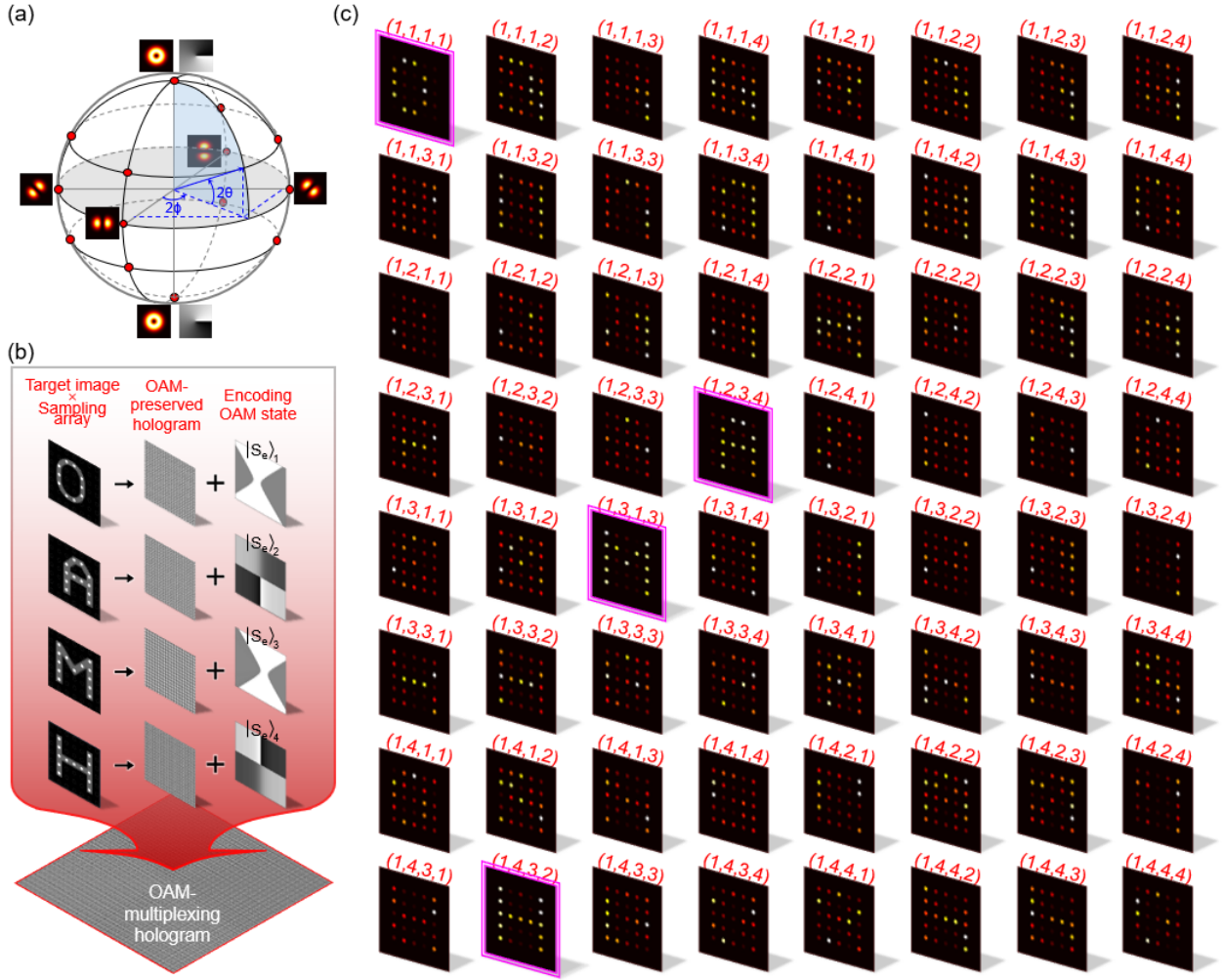

**FIG. 4.** Entanglement-enabled holography for high-security holographic imaging encryption. (a) Poincaré sphere composed of OAM states $|1\rangle$ and $|-1\rangle$. Its north and south poles represent the $|1\rangle$ and $|-1\rangle$. Any superposition state of $|1\rangle$ and $|-1\rangle$ can be represented by a point on the Poincaré sphere. (b) Design of multiplexing hologram based on OAM superposition states and its corresponding OAM-multiplexing hologram displaying on SLM-B. (c) Experimental OAM-selective holographic images for different decoding operators implemented with SLM-A.

In conclusion, we have presented a way to realize the high-dimensional quantum holography based on high-dimensional OAM entangled states. Because different OAM eigenstates are orthogonal to each other, each OAM state can be implemented as an independent information carrier in quantum holography. Giving full play to the high-dimensional characteristics of OAM states, we can obtain a high capacity quantum holographic system by multiplexing a wide range of OAM-dependent holographic images. Proof-of-principle experiment with six-dimensional OAM entangled states has been carried out and verified the feasibility of the high-dimensional OAM quantum holography and high capacity quantum multiplex holography. With the improvement of the generation technology of higher dimensional OAM entanglement states [46,47], higher dimensional quantum holography can be carried out with our method. Comparing with classical holography, quantum holography has practical advantages. The introduction of entanglement can effectively reduce the impact of classical noise, which leads to that the decryption in the presence of strong noise has been achieved. In the quantum OAM holography, the use of superposition state leads to that more states can be selected for encoding information. Therefore, the level of security can be greatly improved in comparison with the classical one. At present, the practical limitation of our quantum holography is its long measurement time, which is taken by scanning every pixel in the measurement process. Luckily, faster and cheaper sensors for imaging quantum correlations are rapid developed. All these make it possible to applied our quantum holographic imaging encryption scheme into practical applications in the near future.

## ACKNOWLEDGEMENTS

This work was supported by the National key R & D Program of China under Grant No. 2022YFA1404900, National Natural Science Foundation of China (12004038 and 11904022).

# Supporting Materials for

# High-dimensional entanglement-enabled holography for quantum encryption

Ling-Jun Kong*, Yifan Sun*, Furong Zhang, Jingfeng Zhang, and Xiangdong Zhang$

Key Laboratory of advanced optoelectronic quantum architecture and measurements of Ministry of Education, Beijing Key Laboratory of Nanophotonics & Ultrafine Optoelectronic Systems, School of Physics, Beijing Institute of Technology, 100081 Beijing, China.

*These authors contributed equally to this work. $Author to whom any correspondence should be addressed. E-mail: zhangxd@bit.edu.cn

## Sec. 1. Experimental data for certifying high-dimensional OAM entangled source.

To certify high-dimensional OAM entangled source, we measure the two-photon coincidence counts between OAM states $\{|l\rangle_A, |l\rangle_B\}$ (for $l = -3, -2, -1, 0, 1, 2, 3$, the experimental results are shown in Figure S1a). These results show that there is still a high coincidence count rate for OAM state $|\pm 3\rangle$. Based on these experimental results, the spiral bandwidth of the source can be estimated with methods used in [1,2]. Furthermore, we also measure the two-photon coincidence counts with the first mutually unbiased basis, which are defined as followings,

$$|M_j\rangle = \frac{1}{\sqrt{d}} \sum_{m=1}^{d} exp\left[\frac{i2\pi}{d}(j-1)(m-1)\right]|l_m\rangle$$

For the 7-dimensional case, $d = 7$, and the experimental results are shown in Figure S1b, which show that the coincidence count rate on the diagonal is relatively uniform and the contrast is also very good.

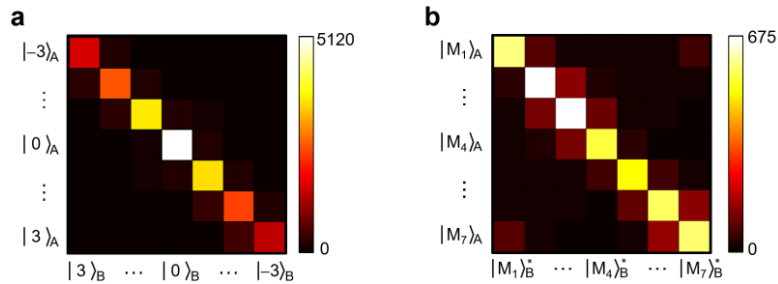

**Figure S1.** Experimental data certifying 7-dimensional entanglement. Two-photon coincidence counts between OAM states $\{|l\rangle_A, |l\rangle_B\}$ (a) and the first mutually unbiased basis $\{|M_j\rangle_A, |M_j\rangle_B\}$ (b).

We also reconstructed the density matrix of the $d$-dimensional OAM entangled quantum states for $d = 5$ and $d = 6$, respectively. For $d = 5$, the projection basis vectors are defined as $|V_1\rangle = |-2\rangle$, $|V_2\rangle = |-1\rangle$, $|V_3\rangle = |0\rangle$, $|V_4\rangle = |1\rangle$, $|V_5\rangle = |2\rangle$, $|V_6\rangle \propto |V_1\rangle + |V_2\rangle$, $|V_7\rangle \propto |V_1\rangle + i|V_2\rangle$, $|V_8\rangle \propto |V_1\rangle + |V_3\rangle$, $|V_9\rangle \propto |V_1\rangle + i|V_3\rangle$, $|V_{10}\rangle \propto$

$|V_1\rangle + |V_4\rangle$, $|V_{11}\rangle \propto |V_1\rangle + i|V_4\rangle$, $|V_{12}\rangle \propto |V_1\rangle + |V_5\rangle$, $|V_{13}\rangle \propto |V_1\rangle + i|V_5\rangle$, $|V_{14}\rangle \propto |V_2\rangle + |V_3\rangle$, $|V_{15}\rangle \propto |V_2\rangle + i|V_3\rangle$, $|V_{16}\rangle \propto |V_2\rangle + |V_4\rangle$, $|V_{17}\rangle \propto |V_2\rangle + i|V_4\rangle$, $|V_{18}\rangle \propto |V_2\rangle + |V_5\rangle$, $|V_{19}\rangle \propto |V_2\rangle + i|V_5\rangle$, $|V_{20}\rangle \propto |V_3\rangle + |V_4\rangle$, $|V_{21}\rangle \propto |V_3\rangle + i|V_4\rangle$, $|V_{22}\rangle \propto |V_3\rangle + |V_5\rangle$, $|V_{23}\rangle \propto |V_3\rangle + i|V_5\rangle$, $|V_{24}\rangle \propto |V_4\rangle + |V_5\rangle$, $|V_{25}\rangle \propto |V_4\rangle + i|V_5\rangle$. The two-photon coincidence counts between $\{|V_j\rangle_A\}$ and $\{|V_{j'}\rangle_R\}$ (with $j$, $j' = 1, 2, \ldots, 25$) has been measured and shown in Figure S2a. The real part and image part of the reconstructed density matrix ($\rho_d$) are shown in Figures S2b and S2c, respectively. The fidelity can be calculated with $F = (tr[(\sqrt{\rho_T}\rho_d\sqrt{\rho_T})^{1/2}])^2 = 0.9$. From our experimental results shown in Figures S1a and S2a, it can be seen that the OAM entangled state prepared in our experiment is not the maximum entangled state. Here, we regard the target state as a pure one, adjust the relative amplitude of each OAM basis vector, and set the target state as

$$|\Psi_T^5\rangle \propto 0.75|-2\rangle_A|2\rangle_B + 0.90|-1\rangle_A|1\rangle_B + |0\rangle_A|0\rangle_B + 0.90|1\rangle_A|-1\rangle_B + 0.75|2\rangle_A|-2\rangle_B$$

Similarly, for $d = 6$, $|V_1\rangle = |-3\rangle$, $|V_2\rangle = |-2\rangle$, $|V_3\rangle = |-1\rangle$, $|V_4\rangle = |1\rangle$, $|V_5\rangle = |2\rangle$, $|V_6\rangle = |3\rangle$, …. The two-photon coincidence counts between $\{|V_j\rangle_A\}$ and $\{|V_{j'}\rangle_R\}$ (with $j$, $j' = 1, 2, \ldots, 36$) has been measured and shown in Figure S2d. The real part and image part of the reconstructed density matrix ($\rho_d$) are shown in Figures S2e and S2f, respectively. The fidelity $F = 0.86$ by setting the target state

$$|\Psi_T^6\rangle \propto 0.67|-3\rangle_A|3\rangle_B + 0.82|-2\rangle_A|2\rangle_B + |-1\rangle_A|1\rangle_B + |1\rangle_A|-1\rangle_B + 0.82|2\rangle_A|-2\rangle_B + 0.67|3\rangle_A|-3\rangle_B$$

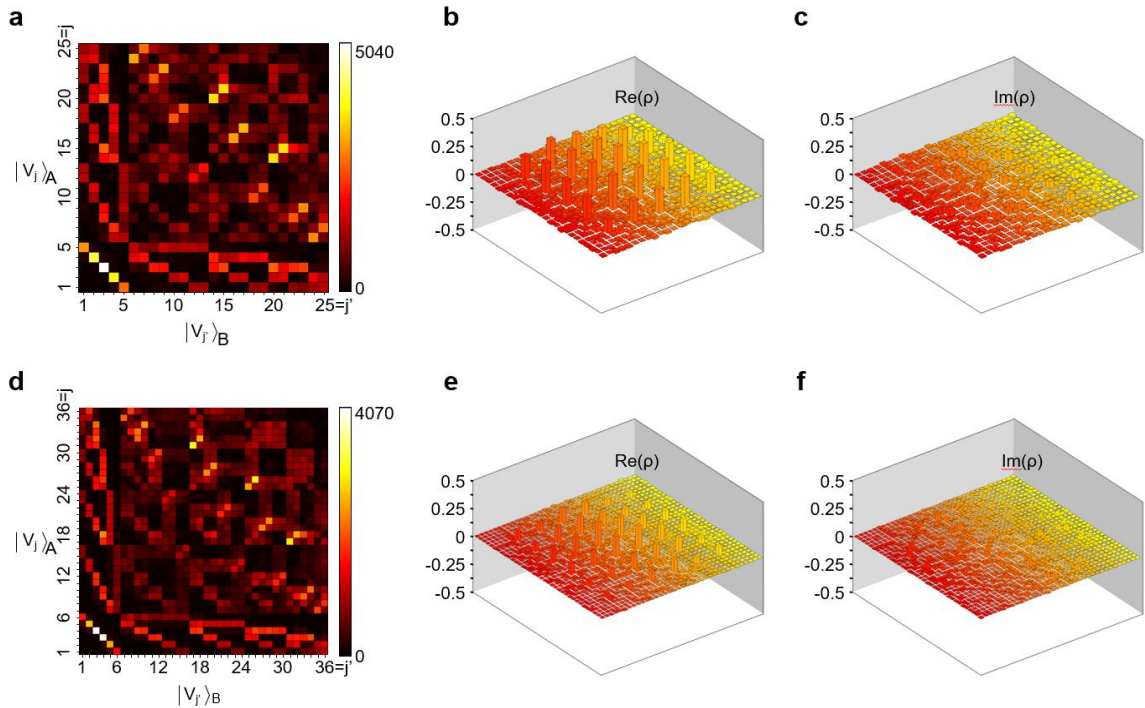

**Figure S2. a**, The two-photon coincidence counts between the complete set $\{|V_j\rangle\}$ for dimension 5. The real part (**b**) and image part (**c**) of the reconstructed density matrix for dimension 5. **d**, The two-photon coincidence counts between the complete set $\{|V_j\rangle\}$ for dimension 6. The real part (**e**) and image part (**f**) of the reconstructed density matrix for dimension 6.

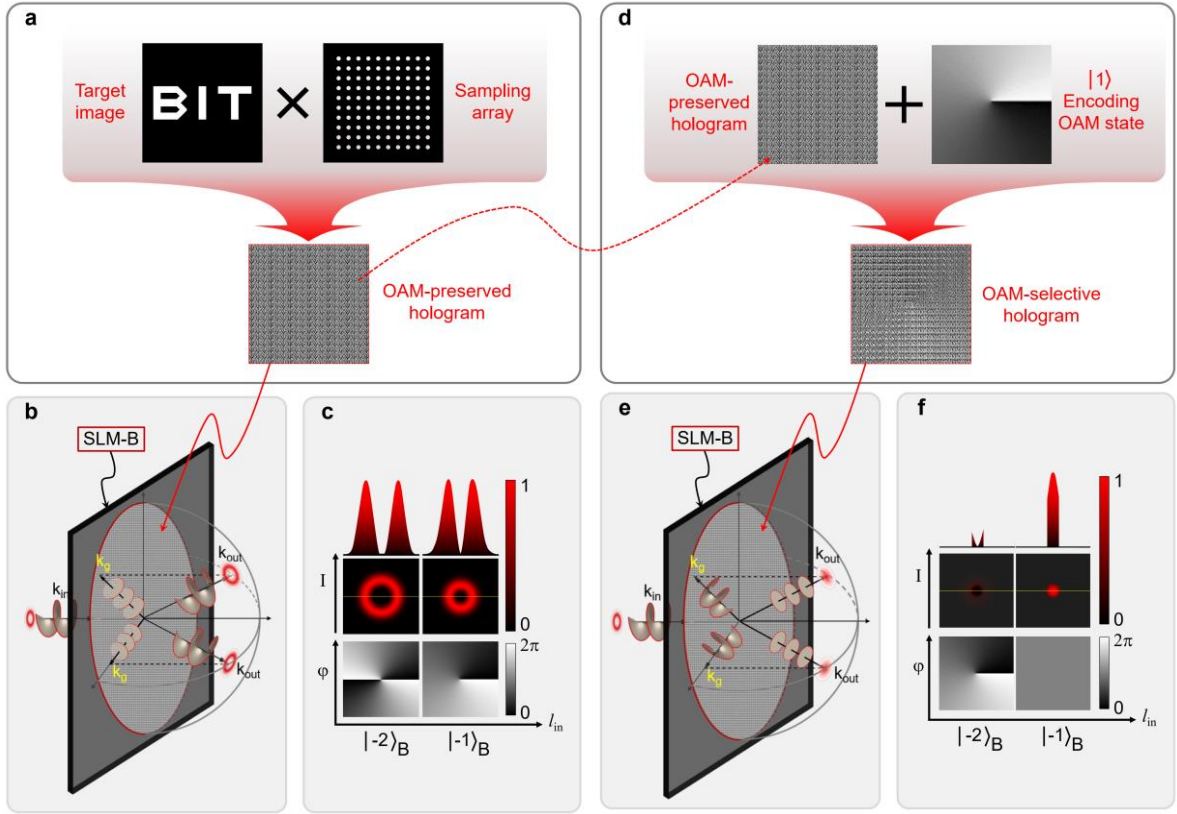

**Figure S3**. Designs of OAM-preserved/selective holograms. **a**, Design of an OAM-preserved hologram by multiplying an object image with a sampling array. **b**, Schematic illustration of the OAM transfer for OAM-preserved hologram in the 3-dimensional spatial frequency domain (k-space). **c**, Intensity (I) and phase ($\varphi$) distributions of single pixels selected from the reconstructed holographic images. **d**, Design of an OAM-selective hologram displaying on SLM-B by adding an OAM state $|1\rangle$ onto an OAM-preserved hologram. **e**, Schematic illustration of the OAM conversion for OAM-selective hologram in the 3-dimensional spatial frequency domain (k-space). **f**, Intensity (I, with only considering the center area) and phase ($\varphi$) distributions of each pixel in the reconstructed holographic images using different OAM modes.

### Sec. 2. OAM-preserved hologram and OAM-selective hologram

In this section, we describe the generations of the OAM-preserved hologram and OAM-selective hologram in details. As shown in Figure S3a, the OAM-preserved hologram is obtained by multiplying a target image (letters 'BIT') with a sampling array, which corresponds to a 2D Dirac comb function. The periodic information of the 2D Dirac comb plays the role as a grating and provide discretely transverse momentums $|k_g\rangle$ to the momentum of input photons $|k_{in}\rangle$ for obtaining discrete momentum of output photons $|k_{out}\rangle$ as shown in Figure S3b. When the OAM-preserved hologram is displayed on the SLM-B in our experiment (shown in Fig. 1(a) in main text), its function for photon-B can be described with the operator $\hat{P}_B^H = e^{j\varphi}|l\rangle_B\langle l|_B$. Here $|l\rangle_B$ denotes a state of photon-B with an OAM of $l\hbar$, $\varphi$ is the phase distribution of the OAM-preserved hologram and contains the information of the target image. In the spatial frequency domain, the function of the operator $\hat{P}_B^H$ should be described as $\hat{P}_B^H|k_{in}\rangle|l\rangle_B = (|k_g\rangle + |k_{in}\rangle)|l\rangle_B = |k_{out}\rangle|l\rangle_B$. Each discrete momentum $|k_{out}\rangle$ will form a pixel on the image plane and the holographic pattern of target image can be reconstructed by these pixels. Clearly, the OAM property of each pixel on the holographic pattern was inherited from the incident OAM states $|l\rangle_B$ directly (Figure S3c).

The OAM selectivity hologram is obtained by encoding OAM state $|l_e\rangle$ onto the OAM-preserved hologram (Figure S3d shown an example with $|l_e = 1\rangle$). In this condition, the operator of SLM-B can be described as $\hat{P}_B^H = e^{j\varphi}|l_e + l\rangle_B \langle l|_B$. While, in the spatial frequency domain, the function of the operator $\hat{P}_B^H$ should be described as $\hat{P}_B^H |k_{in}\rangle|l\rangle_B = (|k_g\rangle + |k_{in}\rangle)|l_e + l\rangle_B = |k_{out}\rangle|l_e + l\rangle_B$. Clearly, the OAM property of each pixel on the holographic pattern was determined by the encoded OAM state $|l_e\rangle$ and the incident OAM states $|l\rangle_B$ together (Figure S3e). Only a given incident OAM state $|-l_e\rangle_B$ can be converted into the Gaussian mode ($|0\rangle_B$) in each pixel of OAM holographic images (Figure S3f).

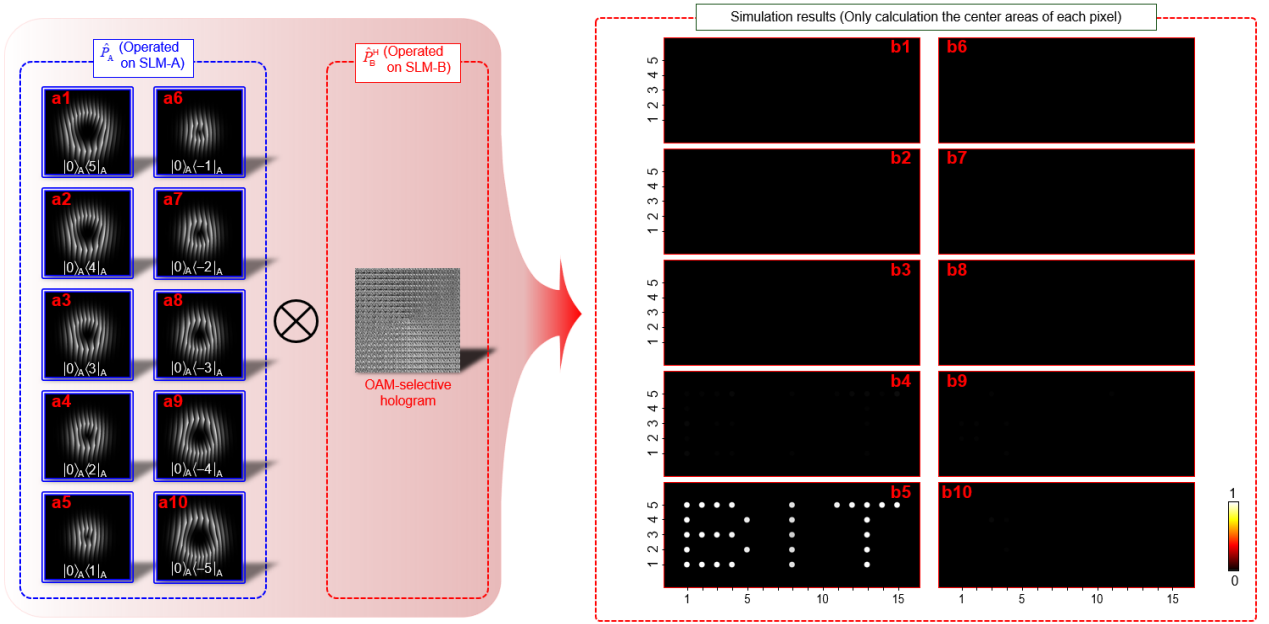

**Figure S4.** Quantum OAM-selective hologram. **a1-a8**, Eight kinds of OAM projective measurement operators of $\hat{P}_A = |0\rangle_A\langle 4|_A$, $|0\rangle_A\langle 3|_A$, $|0\rangle_A\langle 2|_A$, $|0\rangle_A\langle 1|_A$, $|0\rangle_A\langle -1|_A$, $|0\rangle_A\langle -2|_A$, $|0\rangle_A\langle -3|_A$, or $|0\rangle_A\langle -4|_A$ (holograms displaying on the SLM-A). **b***i*, The simulation result of holographic image by only calculating the center areas of 15×5=75 pixels with Eq. (1) in main text for OAM projective measurement operator shown in (**a***i*). Here *i* = 1, 2, 3, 4, 5, 6, 7, or 8. The intensities are normalized to the holographic image reconstructed for the OAM projection with $\hat{P}_A = |0\rangle_A\langle 1|_A$.

## Sec. 3. The simulation of the OAM-selective holography

Figure 1(b) in the main text offers the experimental results of the OAM-selective holography. In this section we provide the corresponding simulation results. In the quantum OAM-selective holography, the OAM-selective hologram displays on SLM-B. Then, the function of the operator $\hat{P}_B^H$ for photon-B is described as $\hat{P}_B^H = e^{j\varphi}|l_e + l\rangle_B\langle l|_B$. The OAM property of each pixel on the holographic pattern was determined by the encoded OAM state $|l_e\rangle$ and the incident OAM states $|l\rangle_B$ together. Only a given incident OAM state $|-l_e\rangle_B$ can be converted into the Gaussian mode in each pixel of OAM holographic images. In other words, only when the OAM projective measurement operator for photon-A satisfies $\hat{P}_A = |0\rangle_A\langle 1|_A$, can each pixel of OAM holographic images be converted into Gaussian mode. Therefore, the OAM-selective holographic image reconstructed (with only detecting the center areas of each pixel) appears (Figure S4b4) only when the projective measurement operator $\hat{P}_A = |0\rangle_A\langle 1|_A$ (Figure S4a4).

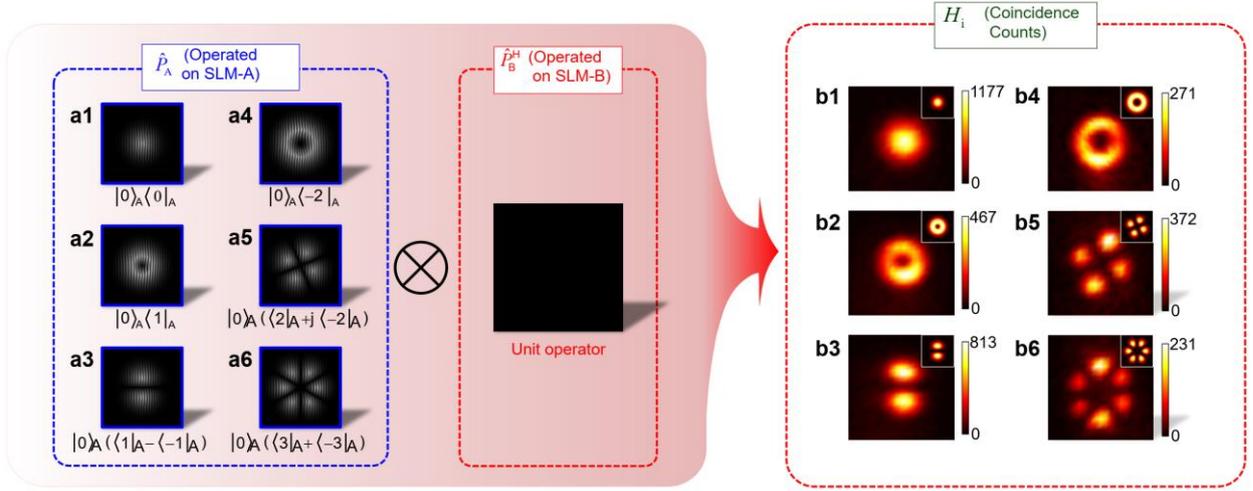

**Figure S5.** Experiment for testing projection operators implemented on SLM-A in our quantum OAM holographic system. Six kinds of OAM projection operators (holograms) displaying on the SLM-A (**a1-a6**) and the corresponding images obtained with coincidence measurements between detectors D-A and D-B (**b1-b6**). During these tests, the phase distribution of the hologram displaying on SLM-B is a constant ($\varphi = constant$), which implies that $\hat{P}_B^H$ is the unit operator.

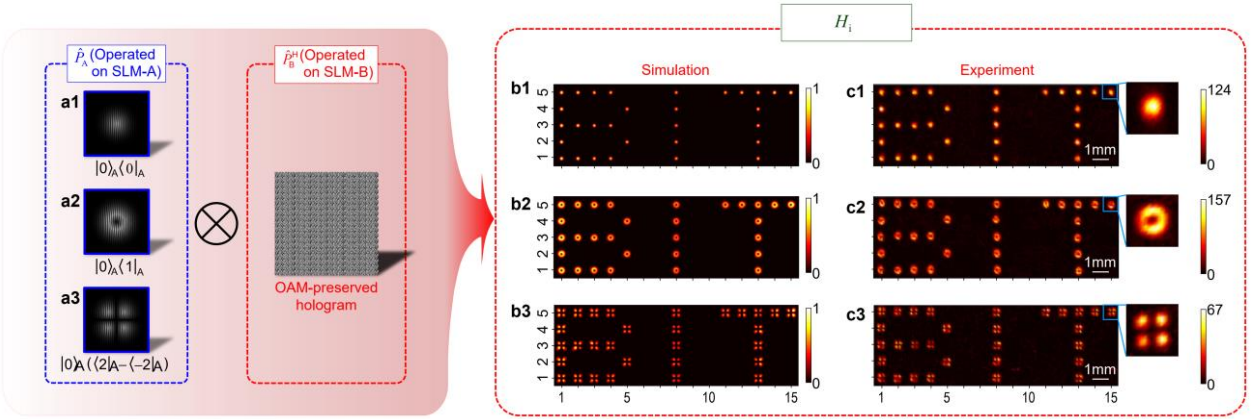

**Figure S6.** Quantum OAM-carrying hologram. **a1-a3**, Three kinds of OAM projective measurement operators of $\hat{P}_A = |0\rangle_A\langle 0|_A$, $|0\rangle_A\langle 1|_A$, or $|0\rangle_A(\langle 2|_A - \langle -2|_A)$ (holograms displaying on the SLM-A). **b1-b3**, **c1-c3**, The corresponding simulation results with Eq. (1) in the main text (**b1-b3**), and the corresponding holographic images obtained with coincidence measurements between D-A and D-B (**c1-c3**). There are 15×5=75 pixels. In experiment, each pixel performs 10×10=100 coincidence measurements and each coincidence measurement takes 10s (**c1**), 20s (**c2**), or 20s (**c3**).

## Sec. 4. Quantum OAM-preserved /selective holography

In this section we provide detailed information on how we carry out the quantum OAM-selective holography step by step. In order to describe the whole process more clearly, we describe our experimental setup first.

**4.1: Testing projective measurement operators implemented on SLM-A in our quantum OAM holographic system.** In this subsection, we show the simulation and experimental results for testing OAM projective measurement operators implemented on SLM-A in our quantum OAM holographic system. During the test, the phase distribution of the hologram displaying on SLM-B is a constant, $\varphi = constant$, which implies that the operator for photon-B is the unit operator $\hat{P}_B^H = I$. OAM projective measurement operators (corresponding to the holograms displaying on SLM-A) can be switched. When $\hat{P}_A = |0\rangle_A\langle 0|_A$ (Figure S5a1), the holographic image can be calculated with Eq. (1) in main text as $H_i = |f\{\langle 0|_A|\Psi\rangle\}|^2 \propto |f\{|0\rangle_B\}|^2$, which mean that the coincidence measurement between detectors D-A and D-B is the intensity distribution of Gaussian mode $|0\rangle$ (Figure S5b1). When $\hat{P}_A = |0\rangle_A\langle 1|_A$ (Figure S5a2), $H_i = |f\{\langle 1|_A|\Psi\rangle\}|^2 \propto |f\{|-1\rangle_B\}|^2$, which mean that the result is the doughnut-shaped intensity distribution of the OAM state $|-1\rangle$ (Figure S5b2). Figures S5(a1-a6) are the OAM projective measurement operators for photon-A with $\hat{P}_A = |0\rangle_A\langle 0|_A$ (a1), $|0\rangle_A\langle 1|_A$ (a2), $|0\rangle_A(\langle 1|_A - \langle -1|_A)$ (a3), $|0\rangle_A\langle 2|_A$ (a4), $|0\rangle_A(\langle 2|_A + j\langle -2|_A)$ (a5), and $|0\rangle_A(\langle 3|_A + \langle -3|_A)$ (a6). The images obtained with coincidence measurements between D-A and D-B are shown in Figures S5(b1-b6), respectively. The corresponding simulation results (shown in insets) calculated with Eq. (1) in main text agree with the corresponding experimental results.

**4.2: Quantum OAM-preserved holography.** In this subsection, we show some calculations and experiments of quantum OAM-carrying holography and Quantum OAM-selective holography with our quantum OAM holographic system. In the quantum OAM-carrying holography, the OAM-preserved hologram displays on SLM-B. In this condition, the function of the operator $\hat{P}_B^H$ for photon-B is described as $\hat{P}_B^H = e^{j\varphi}|l\rangle_B\langle l|_B$. The OAM-carrying holographic image can be calculated with Eq. (1) in main text. When $\hat{P}_A = |0\rangle_A\langle 0|_A$, $|0\rangle_A\langle 1|_A$, or $|0\rangle_A(\langle 2|_A - \langle -2|_A)$ (Figures S6a1-a3), the simulation and experimental results of OAM-carrying holographic images are shown in Figures S6b1-b3, and S6c1-c3, respectively. These results suggest that the OAM property was inherited from the operator $\hat{P}_A|\Psi\rangle$. The success of these quantum OAM-preserved holography has prepared for the progress of the quantum OAM-selective holography.

**4.3: Quantum OAM-selective holography.** In the quantum OAM-selective holography, the OAM-selective hologram displays on SLM-B. Then, the function of the operator $\hat{P}_B^H$ for photon-B is described as $\hat{P}_B^H = e^{j\varphi}|l_e + l\rangle_B\langle l|_B$. Here $\varphi$ is still the phase distribution of the OAM-preserved hologram. The OAM property of each pixel on the holographic pattern was determined by the encoded OAM state $|l_e\rangle$ and the incident OAM states $|l\rangle_B$ together. Only a given incident OAM state $|-l_e\rangle_B$ can be converted into the

Gaussian mode in each pixel of OAM holographic images. In other words, only when the OAM projective measurement operator for photon-A satisfies $\hat{P}_A = |0\rangle_A \langle 1|_A$ (Figure S7a2), can each pixel of OAM holographic images be converted into Gaussian mode (Figure S7b2). Therefore, the OAM-selective holographic image reconstructed with only detecting the center areas of each pixel appear only when the projective measurement operator of $\hat{P}_A = |0\rangle_A \langle 1|_A$.

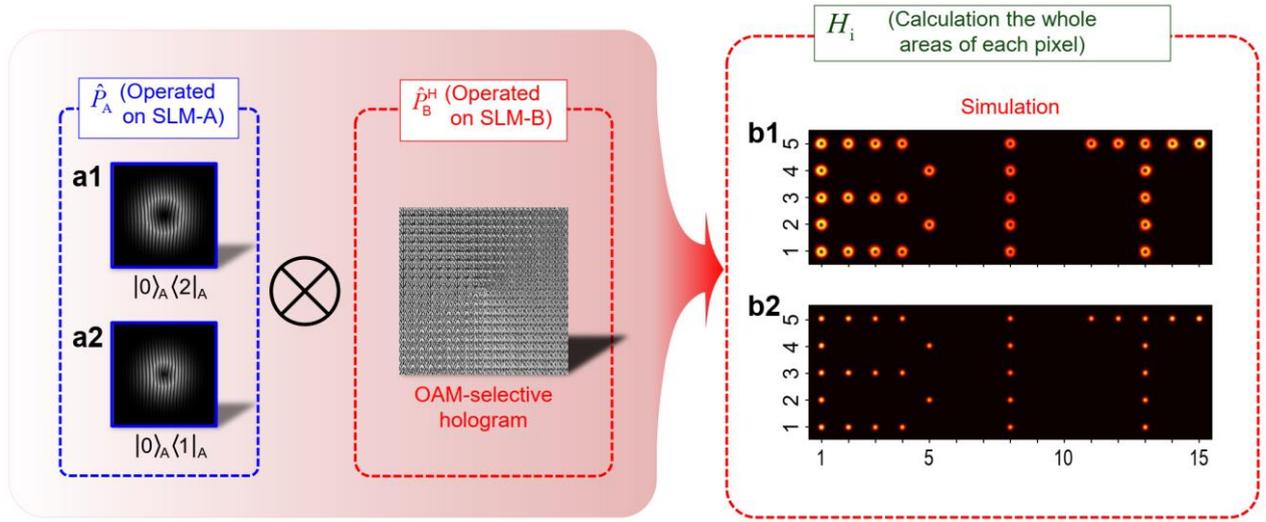

**Figure S7.** Quantum OAM-selective hologram with detecting the whole areas of each pixel. **a1**, **a2**, Two kinds of OAM projective measurement operators of $\hat{P}_A = |0\rangle_A \langle 2|_A$ or $|0\rangle_A \langle 1|_A$ (holograms displaying on the SLM-A), respectively. **b1**, **b2**, The corresponding simulation results of holographic images by calculating the whole areas of 15×5=75 pixels with Eq. (1) in main text.

**4.4: Quantum OAM-multiplexing holography.** In the main text, we give the general theory of quantum OAM-multiplexing holography (as described in Fig. 2(a) in the main text). Here, an example is taken to demonstrate the quantum OAM-multiplexing holography in theory. As shown in Figure S8a, ten letters of the word "HOLoGRAPhY" are the target images. The OAM-preserved hologram of each target image is generated by multiplying it with a sampling array. Then, the OAM-selective holograms are obtained by encoding OAM states $|1\rangle$, $|-1\rangle$, $|2\rangle$, $|-2\rangle$, $|3\rangle$, $|-3\rangle$, $|4\rangle$, $|-4\rangle$, $|5\rangle$, and $|-5\rangle$ onto the OAM-preserved holograms. Finally, the OAM-multiplexing hologram is obtained by combine all OAM-selective holograms.

In the process of image reconstruction, the OAM-multiplexing hologram is displayed on the SLM-B, and each of target image can be reconstructed by using the OAM decoding operator of $\hat{P}_A = |0\rangle_A \langle 1|_A$, $|0\rangle_A \langle -1|_A$, $|0\rangle_A \langle 2|_A$, $|0\rangle_A \langle -2|_A$, $|0\rangle_A \langle 3|_A$, $|0\rangle_A \langle -3|_A$, $|0\rangle_A \langle 4|_A$, $|0\rangle_A \langle -4|_A$, $|0\rangle_A \langle 5|_A$, or $|0\rangle_A \langle -5|_A$ on the SLM-A. The simulation results are shown in Figure S8b.

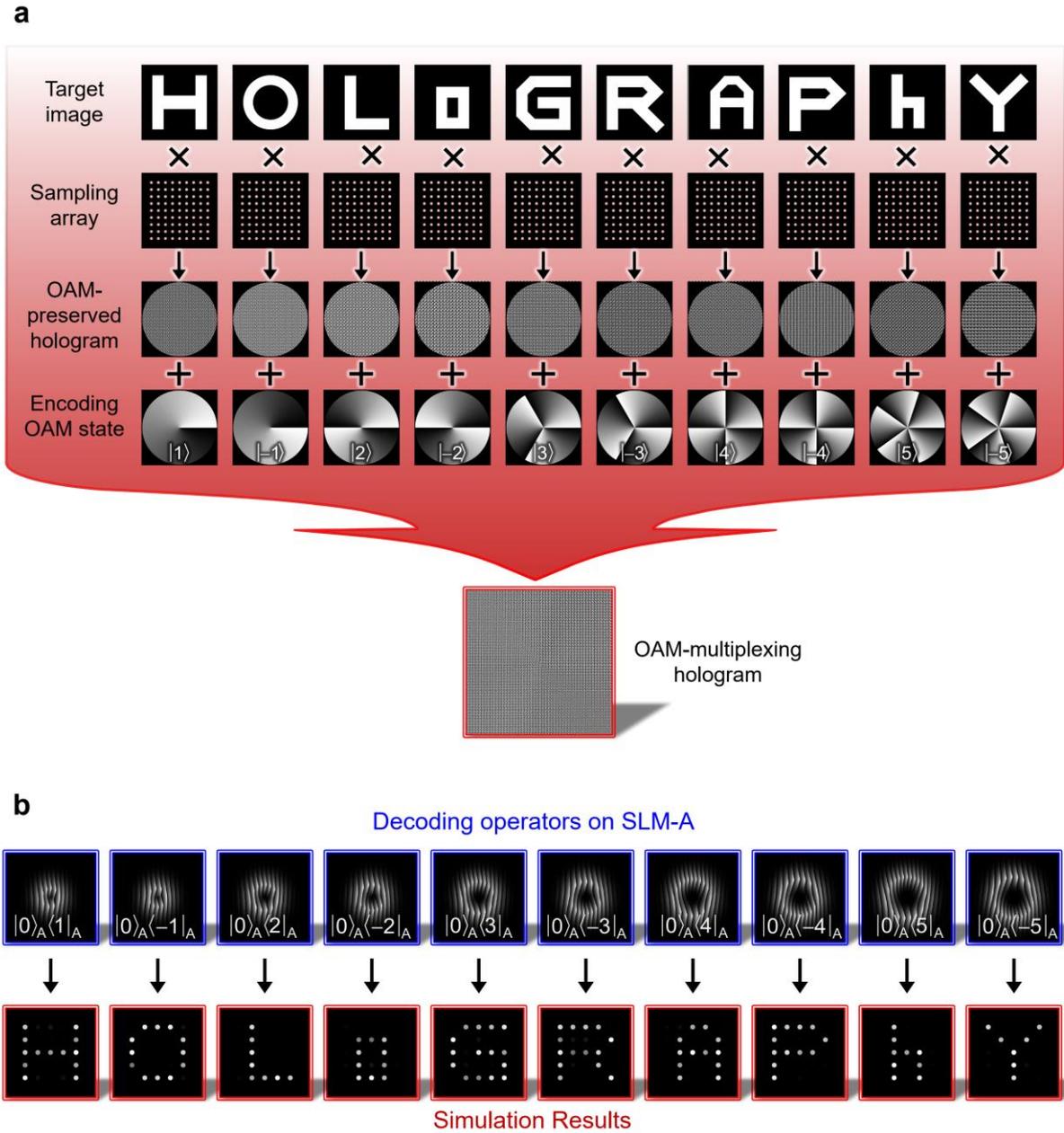

**Figure S8.** Quantum OAM-selective hologram with detecting the whole areas of each pixel. **a**, Generation of an OAM-multiplexing hologram of target images, ten letters of the word "HOLoGRAPhY". **b**, Simulation results of reconstructed target images.

**Sec. 5. The experimental setup for demonstrating the Robustness of our Quantum OAM holography.**
**5.1: Experimental layout of Quantum OAM holography.** As shown in Fig. 1(a) in main text, the pump light is a femtosecond pulsed laser polarized at 0° with a fundamental Gaussian mode, a power of ~600mW, a central wavelength of 405 nm, a pulse duration of ~140 fs, and a repetition rate of 80 MHz (Coherent chameleon). A β-barium borate (BBO) crystal with dimensions of 0.6×7×7 mm$^3$ were cut for producing down-converted photons at a degenerate wavelength of 810 nm in a ~3.0° half-opening angle cone under the type I phase-matching condition. A 650-nm-cutoff long-pass filter was used to block pump photons after the crystals (not shown in main text). There is an interference filter (IF) centred at 810 ± 3 nm in front of each single photon detectors (Perkin Elmer). The SLM was a phase-only modulator (Holoeye Pluto-2-NIR-015) with 1920×1080 pixels and a pixel pitch of 8μm. The focal length of lenses are as follows: L1, 200mm; L2, 500mm; $Lc_1$, 200mm; $Lc_2$, 11mm; Lf, 150mm. The distances between components were as follows: crystal plane to lens L1, 200 mm; lens L1 to Mirror, 200 mm; Mirror to lens L2, 500 mm; lens L2 to SLM, 500 mm; SLM-A to lens $Lc_1$, 200 mm; SLM-B to lens Lf, 150 mm; lens Lf to Fourier plane, 150 mm. The distance between $Lc_1$ and $Lc_2$ is dependent on the distance between $Lc_2$ and the input end of SMF for make sure that SLM-A is imaged onto the input end of SMF. Due to that the distance from $Lc_2$ to the input end of SMF too hard to be measured, the key point is to make sure that when we incident a beam of light from the output end of SMF reversely, light emitted from input end of SMF will be focus on the back focal plane of $Lc_1$ after passing $Lc_2$.

**5.2: Experimental apparatus of OAM-selective holography in the presence of classical noise are shown in Figure S9**. The experimental apparatus of quantum OAM holography is shown in Figure S9a, while the experimental apparatus of classical OAM holographic system is shown in Figure S9b. Their experimental results are shown in Fig. 3 in main text.

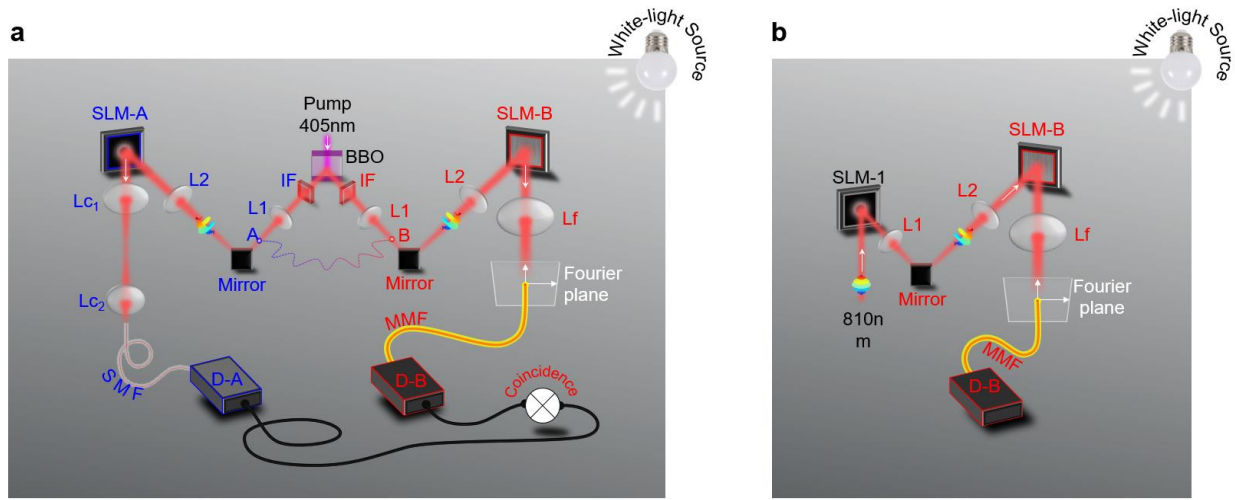

**Figure S9. Experimental apparatus of OAM-selective holography in the presence of classical noise. a**, Experimental apparatus of quantum OAM holography. Comparing with the experimental setup shown in Fig. 1(a) in main text, a white light of a lamp has been added for introducing the classical noise. **b**, Experimental apparatus of classical OAM holographic system. A collimated light beam (810 nm) illuminates SLM-1 for generating incident OAM state. Lenses L1 and L2 image SLM-1 on SLM-B. OAM-selective holographic images measured with recording the counts of D-B in the presence of dynamic stray light for different incident OAM states. For convenient of comparison, SLM-B displays the same OAM-selective hologram used in Fig. 1 in main text in experiment for both **a** and **b**.

**5.3: How to use an MMF as a spatial mode filter.** For comparison, we firstly explain how to realize spatial mode filtering with SMF (which is known very well) [3–5]. In those condition, the focal length of coupling lenses is about 11mm. The intensity distribution of OAM state $|1\rangle$ look like a doughnut, whose hollow diameter is about 5μm (Figure S10a). This is why a single-mode fiber (SMF), whose core diameter is about 5μm, is used. While, in our experiment, the focal length of the lens used to focus photon-B is 150 mm (Fig. 1(a) in main text). The intensity distribution of OAM state $|1\rangle$ look like a doughnut with hollow diameter ~25μm (Figure S10b). This is why a multimode fiber (MMF), whose core diameter is about 25μm, is used as a spatial mode filter.

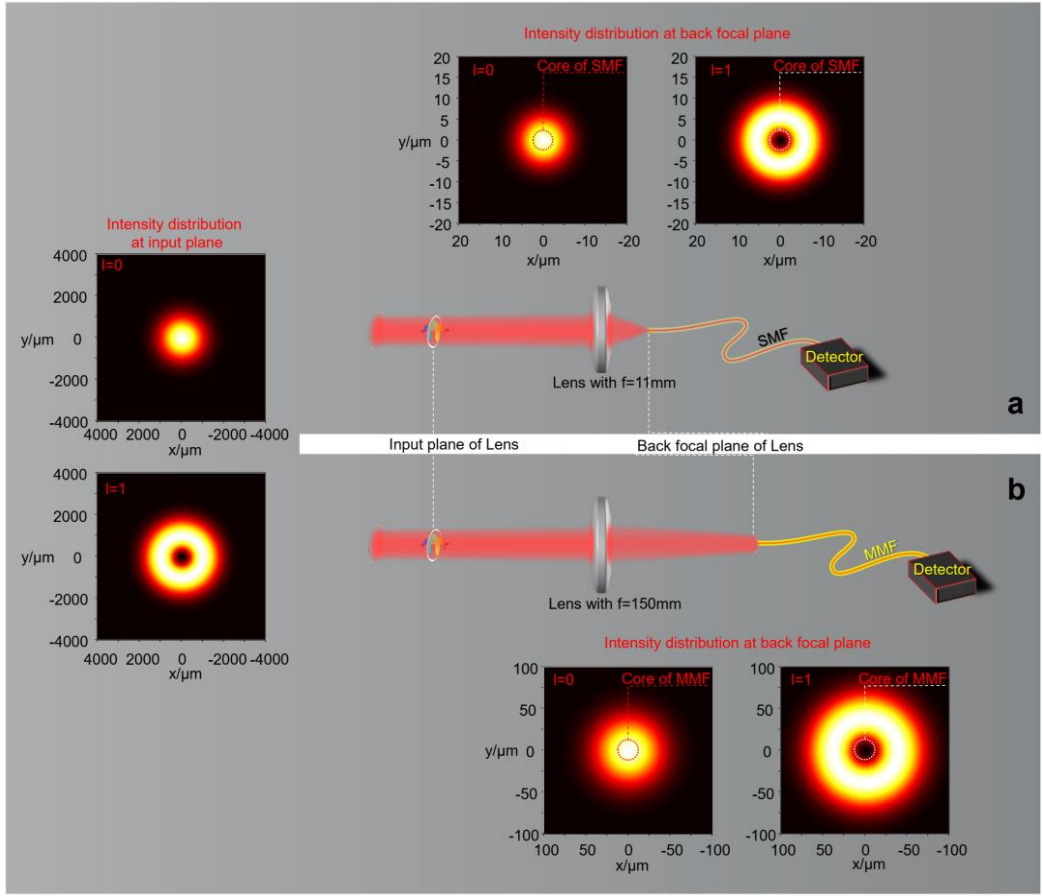

**Figure S10.** spatial mode filtering with SMF (**a**) and MMF (**b**).

**5.4: SNR.** The SNR is defined as $10\log_{10}(C_{Sig}/C_{Noise})$, where $C_{Sig}$ and $C_{Noise}$ represent the average coincidence counts of the desired and undesired pixels of OAM-selective holographic image, respectively. It is easy to obtain that when there is no signal, $C_{Sig} = C_{Noise}$, and thus SNR = 0; when there is no noise, $C_{Sig} = C_{Noise}$, and thus SNR = $\infty$. The value of SNR is affected by pump power, coincidence counting time, pixel number and so on. In our experiment, the pump power (~600 mW) and coincidence counting time (15s for each pixel) are fixed. In the presence of classical noise, the average coincidence counts of both the desired and undesired pixels will be increased by $C_{CN}$. Thus, the SNR decreases to be $10\log_{10}[(C_{Sig}+C_{CN})/(C_{Noise}+C_{CN})]$. For classical OAM holographic system (Figure S9b), the $C_{CN} \gg C_{Sig} \gg C_{Noise}$, and thus SNR~0, which implies that the images will become too vague to tell the letters. While, for quantum OAM holographic system (Figure

S9a), the coincidence measurement decreases the $C_{CN}$ greatly. Leading to $C_{Noise} < C_{CN} < C_{Sig}$, and thus SNR>3, which means that the images is still clear.

**Sec. 6. High-dimensional entangled state for improving the security of quantum OAM holographic encryption in practical application**

As described in the main text, the introduction of the superposition states in the holographic encryption can improve its security due to that the number of superposition states used as information carriers is infinite in principle.

**6.1: The number of distinguishable superposition states in experiment for 2-dimensional case.** As discussed in main text, the level of security of our quantum holographic encryption scheme is improved because the number of information channels (that can be selected for encoding information) is increased by using superposition states as information carriers. The more superposition states, the better the level of security. However, due to the imperfections of preparation, operation and measurement of states in experiment, the overlap of the two states should not be too large to ensure that they can be accurately distinguished, which lead to that the number of superposition states for encoding the information are limited. Here we introduce a parameter ($O$) to characterize how much do the two states overlap. Overlap of states $|S_1\rangle$ and $|S_2\rangle$ is defined as $O = |\langle S_1|S_2\rangle|$. Clearly, $O = 1$ for $|S_1\rangle = |S_2\rangle$ and $O = 0$ for $|S_1\rangle \perp |S_2\rangle$. Next, we discuss the relationship between the number of superposition states ($N_2$) and the value of $O$.

Firstly, we take 2-dimensional condition as an example. As described in the main text, any state can be represented by a point on the Poincaré sphere. Clearly, as the distance between two points on the Poincaré sphere decreases, the overlap of their corresponding states will increases (as shown in the insets of Figure S11a). When listing the discrete superposition states as $\{S\} = \{|S_1\rangle, |S_2\rangle, |S_3\rangle, \ldots, |S_m\rangle, \ldots, |S_n\rangle, \ldots, |S_{N_2}\rangle\}$ and setting the largest overlap between two states in $\{S\}$ to be $O_{max}$, the relationship between the $O_{max}$ and the number of superposition states $N_2$ are shown in Figure S11a. The insets show the distributions of the superposition states (red balls) on the Poincaré sphere (as shown in Fig. 4(a) in main text) for $O_{max} = 0.75$, 0.85, 0.95, 0.97, and 0.98 respectively.

In experiment, the best fidelity of 2-dimensional OAM quantum state is about 0.96 [6]. For pure states, the overlap ($O_{n,m}$) of two state ($|S_m\rangle$ and $|S_n\rangle$) and their fidelity ($F_{n,m}$) satisfies [7]

$$F_{n,m} = \left[Tr\sqrt{\sqrt{|S_n\rangle\langle S_n|}|S_m\rangle\langle S_m|\sqrt{|S_n\rangle\langle S_n|}}\right]^2$$
$$= \left[Tr\sqrt{|S_n\rangle\langle S_n||S_m\rangle\langle S_m||S_n\rangle\langle S_n|}\right]^2$$
$$= |\langle S_n|S_m\rangle|^2 \left[Tr\sqrt{|S_n\rangle\langle S_n|}\right]^2 = O_{n,m}^2$$

Therefore, and the number of corresponding superposition states is about 80 (as shown in Figure S11a), which is so limited that it is difficult to ensure the security of our quantum OAM holographic encryption system. Fortunately, such a problem can be overcome by using high-dimensional entangled states, because the number of the distinguishable superposition states increases rapidly with the increase of dimension.

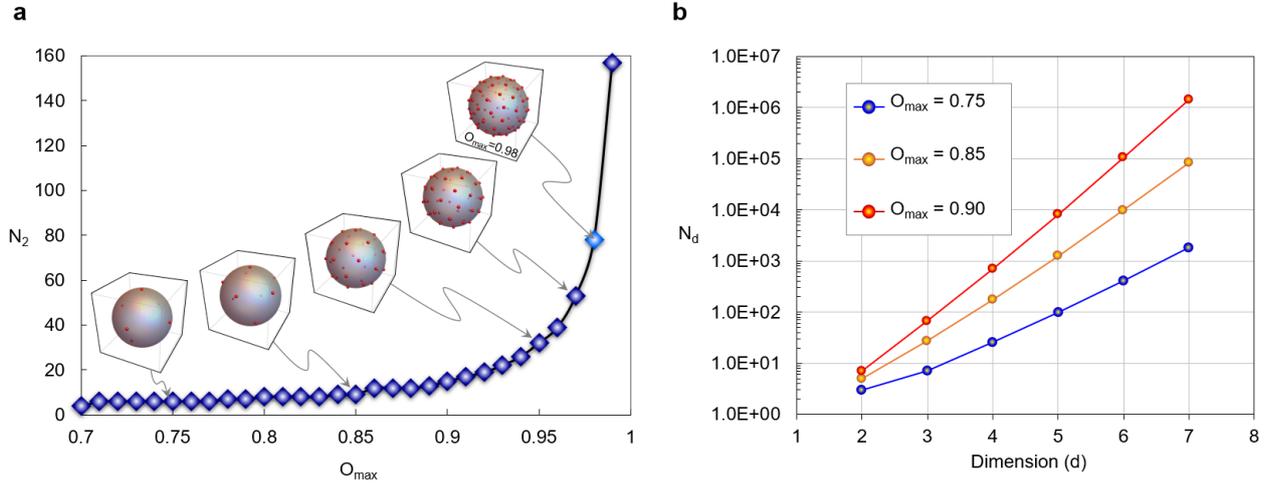

**Figure S11. a**, The relationship between the largest overlap ($O_{n,m}$) and the maximum number of the discrete and distinguishable superposition states ($N_2$) for 2-dimensional case. **b**, The relationship between the $N_d$ and the dimension for $O_{max}$ =0.75, 0.85, and 0.90, respectively.

**6.2: The number of distinguishable superposition states in experiment for high-dimensional cases.** Similarly, for high-dimensional condition, the relationship between the $O_{max}$ and the number of superposition states $N_d$ can also be calculated. By setting the discrete superposition states as $\{S\}_d = \{|S_1\rangle, |S_2\rangle, |S_3\rangle, \ldots, |S_m\rangle, \ldots, |S_n\rangle, \ldots, |S_{N_d}\rangle\}$ and the largest overlap between two states in $\{S\}_d$ to be $O_{max}$, the relationship between the number of distinguishable superposition states ($N_d$) and the dimension ($d$) for $O_{max}$ =0.75, 0.85, and 0.90 are shown in Figure S11b. Clearly, the $N_d$ will increase rapidly with the increase of dimension $d$. For 7-dimensional OAM quantum state with $O_{max} = 0.90$ (achievable in experiment [4]), the number of distinguishable superposition states will be about $10^6$, which should be enough to ensure our quantum holographic encryption system has enough security level.